# Quantum key establishment via a multimode fiber

LYUBOV V. AMITONOVA,[1,2,3,*] TRISTAN B. H. TENTRUP,[1] IVO M. VELLEKOOP,[4] AND PEPIJN W. H. PINKSE[1]

[1]*Complex Photonic Systems (COPS), MESA+ Institute for Nanotechnology, University of Twente, PO Box 217, 7500 AE Enschede, The Netherlands*
[2]*LaserLaB, Department of Physics and Astronomy, Vrije Universiteit Amsterdam, De Boelelaan 1081, 1081 HV Amsterdam, The Netherlands*
[3]*Current address: Advanced Research Center for Nanolithography, Science Park 106, 1098 XG Amsterdam, The Netherlands*
[4]*Biomedical Photonic Imaging, TechMed Institute, University of Twente, PO Box 217, 7500 AE Enschede, The Netherlands*
**l.amitonova@arcnl.nl*

**Abstract:** Quantum communication aims to provide absolutely secure transmission of secret information. State-of-the-art methods encode symbols into single photons or coherent light with much less than one photon on average. For long distance communication, typically a single-mode fiber is used and significant effort has been devoted already to increase the data carrying capacity of a single optical line. Here we propose and demonstrate a fundamentally new concept for remote key establishment. Our method allows high-dimensional alphabets using spatial degrees of freedom by transmitting information through a light-scrambling multimode fiber and exploiting the no-cloning theorem. Eavesdropper attacks can be detected without using randomly switched mutually unbiased bases. We prove the security against a common class of intercept-resend and beam-splitting attacks with single-photon Fock states and with weak coherent light. Since it is optical fiber based, our method allows to naturally extend secure communication to larger distances. We experimentally demonstrate this new type of key exchange method by encoding information into a few-photon light pulse decomposed over guided modes of an easily available multimode fiber.

## 1. INTRODUCTION

The importance of secure communication is rapidly growing [1]. We use cryptography in everyday life often without noticing, for example, when we conduct financial transactions via the internet. The security of conventional cryptography is based on shared secret keys or on computational assumptions, such as the presumed hardness of factoring [2]. In practice, this means that it is vulnerable to unanticipated advances in hardware or algorithms. Quantum cryptography in theory provides unconditional secure communication, assuming only that an eavesdropper (Eve) is restricted by the laws of physics: the quantum no-cloning theorem forbids to replicate an unknown quantum state [3]. Indeed, the security of quantum cryptography requires quantum states of light [4]. For example, in the original and best-known quantum key distribution (QKD) method – BB84 proposed by Bennett and Brassard [5] – the security is based on the fact that the polarization of a single photon can be prepared and measured along well-defined directions. Key-distribution methods do not by themselves communicate useful information, but such a communication can follow with the proven-secure one-time-pad method after the secure key is built up.

Nowadays, optical fibers are key elements of worldwide communication networks [6]. Single-mode fibers are widely used to transmit voice, television and internet data. The ultimate goal is to increase the data-carrying capacity of a single line. Combining multiple strands of fiber in a single optical cable enable various spatial division multiplexing schemes. This allows the merging of high-dimensional QKD systems with optical fiber transmission. Multicore fibers are also being studied for space-division multiplexing [7–9] and have been used for high-dimensional QKD [10]. A step-index multimode (MM) fiber supports a significantly higher number of modes than a multicore fiber and as a result can transfer information at a higher density. Orbital angular momentum states of a MM fiber enable high-dimensional quantum communication and were implemented to transmit 4- and 3-dimentional quantum states [11,12]. Another method for secure communication relies on optical reciprocity in MM fibers scrambling wavefronts in a random way [13].

Although short pieces of straight or slightly bent fiber are not truly random [14], in any realistic fiber random bends, index imperfections, and other perturbations cause the signal to couple into multiple modes [15], leading to an arbitrary mixing of field amplitudes [16] and scrambling the information across the modes. However, it is well-known that the mode mixing can be partially undone by applying techniques from complex wavefront shaping, a method originally developed for precise light control through and in highly scattering materials [17–19]. Recently methods have been proposed for high-speed [20], high-resolution [21,22] image transfer. Multimode optical fibers can now also be used to transmit information in the spatial domain [23–25].

Here we propose a new method for secure key establishment via a MM fiber. The idea is based on secure characterization of the multimode transmission channel by means of weak light pulses. As can be seen in Fig. 1, both the sender, Alice, and the receiver, Bob, control a stretch of fiber that is randomly bent so that it spatially scrambles the optical communication signal. Eve has no access to the scrambled parts of Alice and Bob: the configurations for both scrambling fibers effectively serve as the encryption and decryption key. Our method is designed such that:

1) Alice and Bob can characterize the scrambled communication channel in a calibration phase and undo the scrambling using wavefront shaping in the communication phase.
2) Eve cannot extract enough information to identify a single transmitted symbol by a measurement (see security analysis).
3) By merit of the no-cloning theorem, Eve cannot clone the transmitted symbol.
4) Eve cannot average over multiple measurements, since Alice sends uncorrelated, randomly and uniformly picked symbols.
5) Eve cannot determine the configurations of the scramblers, even when she is intercepting the optical signals in the calibration and/or communication phases. Since she doesn't know the configuration of the scramblers, by 2-4), she cannot decode the signal.
6) If Eve tries to hide her interception by resending light to Bob, this can easily be detected. We can quantify the amount of information that she can collect when tapping a signal.

Elements 1-4 and 6 are sufficient to ensure secure key establishment in the case that Eve has no method to build a physical copy of the entire multimode communication channel. Element 5 ensures that our method remains secure in the case that Eve has access to some technology to clone or mimic exactly the stretches of distorted multimode fiber.

Our method does not require the light to be in an entangled (or otherwise special) quantum state. Nowadays, many QKD systems (including commercially available ones) rely on a weak coherent state due to its simplicity. However, only since the introduction of decoy protocols, weak coherent states do not compromise security if less than one photon per pulse is used [26,27]. We prove that our method is secure against a *non-adaptive intercept-resend* and a *photon-number splitting* attacks with weak coherent light even when several photons per symbol are used without raising the complexity of the method by implementing the decoy states. However, the concept of decoy states can be used to further improve the performance of our method.

In contrast to traditional QKD our method doesn't require two mutual unbiased bases, which are conventionally used to detect Eve's attacks. As a result, it also doesn't need a quantum-key-sifting step. It is important to mentioned that some modern QKD protocols, such as coherent one-way (COW) [28,29], also do not rely on mutually unbiased bases for sending data bits and allows to transfer the data in deterministic way. In COW, Alice sends full and empty pulses and Bob temporally distinguishes them; decoy states are used in a conjugate basis for security [30]. In contrast to our proposed approach, COW allows to transmit only one bit of the information per pulse.

## 2. THE METHOD

Firstly, we briefly discuss the main idea of the method under ideal conditions assuming a perfect single-photon source and ideal wavefront shaping (in phase and amplitude), since this is the simplest to understand conceptually. Later, we give a quantitative argument why the method is still secure with weak coherent light and imperfect wavefront shaping without any adaptation to the method or the setup. The experimental setup and illustration of the main principle are presented in Fig. 1. Alice and Bob each have under full control a short fragment of MM fiber, which acts as a 'scrambler'. They both 'program' the scrambler in a random way by applying a random conformational change to the fiber. In the calibration phase, Alice and Bob perform a stepwise sequential wavefront shaping algorithm [31,32]: Alice generates a plane wave with a single segment that is phase shifted. Bob records a camera frame and sends to Alice intensities of $S$ randomly chosen points via a classical channel. This is repeated for each segment several times. Alice can now use wavefront shaping to focus light on a desired position on Bob's camera. In this way, we can encode a high-dimensional alphabet into the guided modes of a multimode fiber. The set of special superpositions of fiber modes that lead to light focused at desired points on Bob's side we will call a basis.

To send a symbol, Alice prepares a single-photon quantum state with an appropriately phase-shaped and amplitude-shaped wavefront that leads to a focus of light at the particular position on the fiber output facet. Throughout the fiber, the single photon will be present as a disordered superposition of almost all fiber modes. If a potential eavesdropper Eve would intercept the photon and determine its position, the photon collapses at a nearly random position on Eve's detector, which varies even between identical copies of the same symbol (see Fig. 1 'Eve'). Therefore, Eve will not be able to identify what symbol was sent: the information is scrambled. Only at Bob's end, the photon will be spatially localized at the target position, where each target position corresponds to a symbol from the alphabet (see Fig. 1 'Bob'). Since here the photon is spatially localized, Bob can unambiguously identify what symbol was sent: the information is *unlocked*. As a result, Bob can decode the information instantaneously during the communication via the multimode fiber. Under ideal conditions, no classical postprocessing is required. In case of nonperfect wavefront shaping and/or losses within the system, Bob reports to Alice which characters should be discarded from the final key, as a last step.

The summary of the protocol is as follows:

- Alice and Bob randomly scramble the optical communication channel and characterize it for $S$ randomly distributed detectors in a secure way by using a low number of photons and the modified sequential wavefront shaping algorithm. Alice checks the mutual interference of all the measured wavefronts to verify the security of the wavefront shaping.
- Alice prepares light in one of the $S$ states by applying a spatial phase profile and sends it to Bob on the quantum channel. Bob measured the intensities at $S$ detectors. This step is repeated many times to generate the raw key.
- Alice and Bob communicate over the classical channel and estimate a raw key. Than they reveal a random sample of the bits of their raw keys and estimate the error rate in the quantum channel.
- If the security is verified by an acceptable low error rate, Alice and Bob generate the secret key with the classical processes of error correction and privacy amplification and obtain a shared secret key.

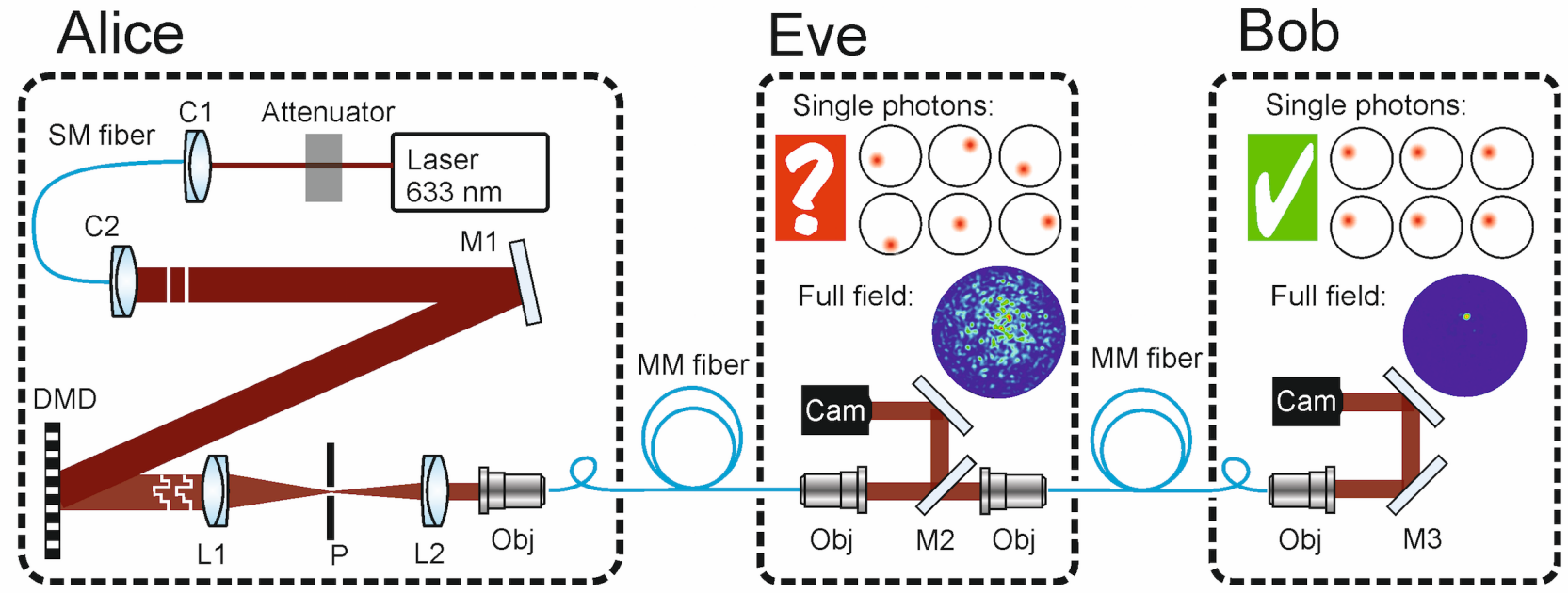

Fig. 1. Experimental setup. Alice encodes a symbol by coupling light with an appropriate phase pattern into a MM fiber. Bob receives a signal focused at a particular position on the fiber output facet (see 'Full field' on Bob's part). Bob can reconstruct the position even with a single detected photon (see 'Single photons' on Bob's part). With mirror M2 Eve can tap off the signal in the middle of the fiber. However, due to the mode-mixing nature of the MM fiber, her intensity distribution is a complex interference pattern (see 'Full field' on Eve's part). The positions of her photon detections vary even for different realizations of the same symbol (see 'Single photons' on Eve's part). As a result, Eve cannot reconstruct the symbol that Alice sent. Abbreviations: DMD, digital micromirror device; SM fiber, single mode fiber; M, mirror; L, lens; Obj, objective; P, pinhole; C, collimator; Cam, camera.

## 3. RESULTS

### *3.1 Experimental demonstration*

We perform our experiments with a weak coherent light source, in the presence of noise, and assuming only a moderate efficiency of wavefront shaping. First, the fiber is calibrated in its current configuration as described in Appendix A. In the experiments, we use a fiber with an approximate number of $N$ = 1500 modes and the alphanumeric 36-dimensional alphabet consisting of A-Z + 0-9 (case insensitive) symbols. The symbols are encoded into $S$ = 36 different positions on the fiber output facet, as presented in faint green circles in Fig. 2(a).

In the first set of experiments, we emulate a single-photon source by taking into account only the frames with single-photon detection events. Snapshots of the spatial distributions of photons measured by Bob and by Eve are presented in Fig. 2(a) and 2(b), respectively. For clarity, we show frames which correspond to the fidelity of wavefront shaping $\alpha^2 = 0.6$ (for the definition of fidelity see Appendix A) and for only two different symbols sent by Alice by open and filled dots. Different colors represent 10 different repetitions of each of the two symbols. The dashed line shows the fiber core edge. In contrast to Eve, Bob clearly sees the correlation between different realizations of the same symbol. In case of a perfect single-photon light source, the probability for Bob to detect the photon in the correct position is equal to the fidelity of wavefront shaping that, as was shown before, can experimentally reach a theoretical maximum of $\pi/4$ for phase-only wavefront shaping [23]. In contrast, different realizations of the same field pattern hardly correlate to each other on Eve's side (see Fig. 2(b)).

In the second set of experiments, we characterize the communication between Alice and Bob for low-fidelity wavefront shaping, $\alpha^2 = 0.1$, and assuming a weak coherent light source with an average number of photons per pulse $\mu^2$ starting from 2. Alice sends each of the 36 symbols 200 times in random order. Each time, Bob reads the signal on the fiber output and estimates what symbol was sent. Bob can use two main strategies. In the first strategy, Bob compares the number of photons in different predefined areas and chooses the one with the highest intensity. In the second strategy, Bob selects only those transmissions for which he is very sure what symbol was sent by accepting only the symbols with at least as many photons as a particular threshold. Whenever Bob has more information than Eve, the few-percent error rate can be corrected down to the standard $10^{-9}$ during the (classical) error correction step of the protocol [33].

We investigate the percentage of the correct reading and the rejection ratio for two different thresholds. Then we repeat the whole procedure for a different average number of photons per pulse. The results averaged over all 36 symbols are presented in Fig. 2(c), where the probability $p$ to detect the correct symbol versus the average number of photons in a pulse, $\mu^2$, is plotted for different strategies of Bob: black, no threshold; red, two-photon threshold and blue, three-photon threshold. Dots represent the experimentally measured data. The black line shows the theoretical prediction from the formula $p = e^{-\lambda_1}\sum_{k=0}^{\infty}[F(k-1)]^{S-1}\lambda_1^k/k!$, where $F(k) = e^{-\lambda_2}\sum_{l=0}^{k}\lambda_2^l/l!$ is the cumulative distribution function of the Poisson distribution, $\lambda_1$ is the average number of photons at the "correct" position, $\lambda_2$ is the average number of photons at the "wrong" position, and $S$ is the total number of symbols. Values $\lambda_1 = 0.1\mu^2$ and $\lambda_2 = 0.005\mu^2$ were extracted from the experimental data. The blue and red lines represent the results of simulations of Bob's probability for a threshold of 2 photons and 3 photons, respectively. Although even in the case of low fidelity of wavefront shaping ($\alpha^2 = 0.1$) and no threshold the probability of reading the correct answer is significant, it dramatically increases for strategies with a threshold. The rejection rates are also carefully analyzed and presented in Fig. 2(d) for a threshold of 2 photons (red bars) and 3 photons (blue bars). The width of the bars corresponds to the standard deviation of the number of photons in a pulse and increases according to the Poisson distribution. As a result, Bob can read information with an error rate close to zero.

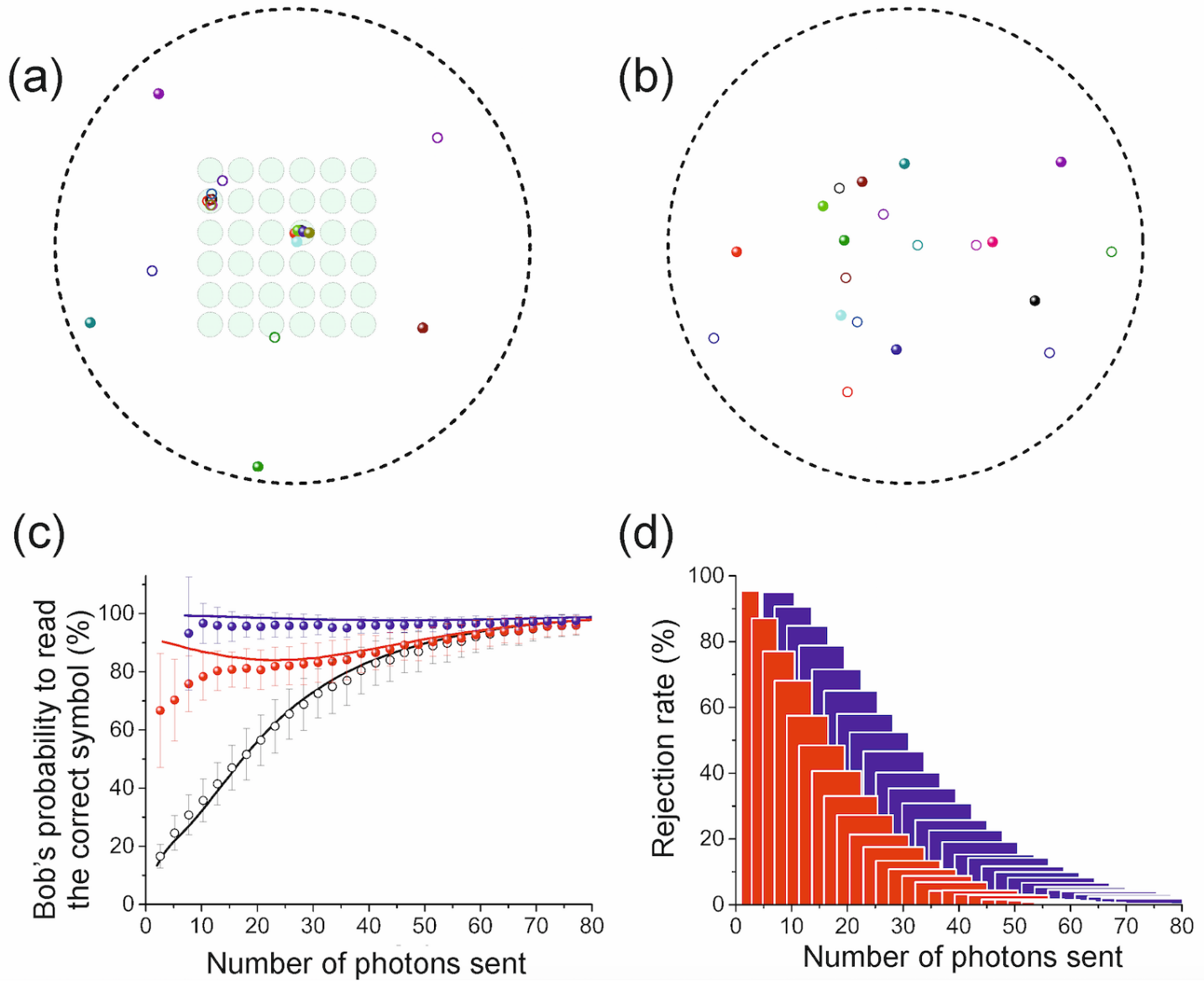

Fig. 2. (a, b) Snapshots of the spatial distributions of photons measured by Bob (a) and by Eve (b) in the case of a perfect single-photon source and a fidelity of wavefront shaping $\alpha^2 = 0.6$. The open and filled dots correspond to two different symbols sent by Alice. Different colors represent 10 different repetitions of each of the symbols. The dashed line indicates the fiber core edge. Faint green circles on Bob's facet represent the symbol areas. In contrast to the photon distribution measured by Eve, Bob clearly sees the correlation between different realizations of the same symbol. (c) The measured probability of Bob to detect the correct symbol versus the average number of photons in a pulse for wavefront shaping fidelity $\alpha^2 = 0.1$, a weak coherent light source and different strategies: Bob uses no threshold (black); Bob accepts only symbols that have been triggered by two or more photons (red) and three or more photons (blue). Vertical error bars represent the standard deviation after averaging over 36 symbols. Solid lines show the theoretical calculations of the probabilities as described in the main text. (d) The rejection rate for a threshold of 2 photons (red bars) and 3 photons (blue bars) versus the average number of detected photons in a pulse. The width of the bars corresponds to the standard deviations of the number of photons in a pulse.

Figure 2(c) demonstrates that our experimental results are consistent with a theoretical analysis for a wavefront shaping fidelity, $\alpha^2 = 0.1$. In the next sections, we will use these results to calculate Bob's probability to read the correct symbol for a different level of wavefront shaping fidelity and number of modes.

### *3.2 Security analysis for a single-photon light source*

Let us assume that the eavesdropper can tap off the signal somewhere in the middle of the fiber. Eve can only retrieve a fraction of the information that was sent by taking snapshots of the complex wavefronts in the fiber, since she cannot sort the wavefronts in $S$ symbols in a passive linear way such as happens at Bob's fiber end. In order to employ classical error correction and privacy amplification, it is necessary that $H_B > H_E$ [33]. In Appendix B, we derive a strict upper limit for the amount of information that Eve can gain in case of *a single-photon light source* as $\langle H_E \rangle = (1 - \gamma)/\ln 2 \approx 0.61$ bit per transmitted symbol regardless of the number of symbols or the number of modes, where $\gamma$ is the Euler constant. Under the same conditions Bob at maximum can have $H_B = \log_2(S) \approx 5.2$ bit of information per transmitted symbol for $S = 36$ symbols. As a result, $\langle H_E \rangle \ll H_B$, guaranteeing the security of the method for a single-photon light source.

### *3.3 Security analysis for weak coherent light*

If rather than single photons *weak coherent light* is used, phase measurements are possible. Here we consider that Eve performs perfect phase measurements. The fundamental limit of the best possible fidelity $\beta^2$ with which Eve can measure the wavefront with a low photon budget is determined by the following expression: $\beta^2 = \mu^2/(\mu^2 + 2N)$ (see Ref. [34] for the details), where $\mu^2$ is the average number of photons per pulse sent by Alice and $N$ the number of fiber-guided modes. The factor 2 appears due to the complicated structure of the fiber-guided modes that mostly include two orthogonal polarizations and consequently requiring two measurements to be done. Figure 3(a) represents Eve's best possible fidelity $\beta^2$ as a function of the number of photons, $\mu^2$, and the number of modes, $N$. We see that even for a high number of photons measured by Eve the fidelity $\beta^2 << 1$. As a result, an intercept-resend attack significantly decreases the proportion of energy in the focus at Bob's side: $\alpha^2 \rightarrow \alpha^2 \beta^2$ resulting in radical rise of the error rate (see Appendix B Eqs. B7-9). For $\beta^2 < 0.01$, an intercept-resend attack on all photons would induce a qudit error rate of more than 0.9.

We now calculate the amount of information per transmitted symbol that Eve can retrieve at best. The example for $N = 1500$ fiber modes and $S = 36$ symbols is presented in Fig. 3(b) where the blue line represents the exact upper limit (evaluated through numerical integration, and the red line represents a simple closed-form analytical upper limit (see Appendix B Eqs. B16-18 for the details). We see that $H_E$ is below the maximum entropy achievable by Bob ($\log_2 S$) even when the average number of photons exceeds unity. We analyze how many photons Eve needs to measure to get the same amount of information as Bob has (Eq. B6). The results are present in Fig. 3(c), showing the maximum number of photons detected by Eve per symbol for which $H_E < H_B$, plotted as a function of the wavefront-shaping fidelity, $\alpha^2$ and the number of modes, $N$ for the number of symbols $S = 36$. Even in the case of a *weak coherent light source* with more than 1 photon per pulse on average Eve is not able to retrieve full information. Note that we have to explicitly exclude the scenario where Eve knows part of the message, since in this case she could sum the measurements for corresponding symbols together, eventually collecting enough information. Therefore, our method is limited to distributing messages which appear completely random to Eve.

The security of the method against intercept-resend attack is based on a robust way to detect the presence of Eve. To estimate the maximum attenuation, we calculate the qudit error rate by using Eqs. B7-9 and the dark count rate of the detectors, $p_{\text{dark}} = 7.2\times10^{-8}$ [35]. The results are presented in Fig. 3(d) where the secure qudit error rate between Alice and Bob is plotted as a function of the total attenuation in dB for $N = 5000$ modes, wavefront shaping fidelity $\alpha^2 = 0.7$, and the average number of emitted photons $\mu^2$ is 0.1 (the black line), 1 (the red line), and 10 (the blue line). An intercept-resend attack on all photons induces an error rate of more than 0.9 at best (the grey dashed line). As a result, the presence of Eve can be easily detected and the protocol is secure even in case of a *weak coherent light source* for more than 40 dB of total attenuation.

One may argue that Eve could, hypothetically, copy and physically reproduce all essential features of the fiber between Eve and Bob in some passive optical system that would focus each symbol onto a different detector. Constructing such a programmable mode sorter which implements any linear transmission matrix is theoretically possible, although it remains an extremely challenging technological problem [36,37]. However, even with the ultimate technology, Eve cannot build a mode sorter without knowledge of the exact transmission matrix. Obviously, Eve cannot directly measure the transmission matrix of the scramblers of Alice and Bob because these parts of the fiber are always kept under the control of Alice and Bob. Therefore, we only have to make sure that Eve cannot find out the transmission matrix by intercepting the calibration phase and/or the communication phase.

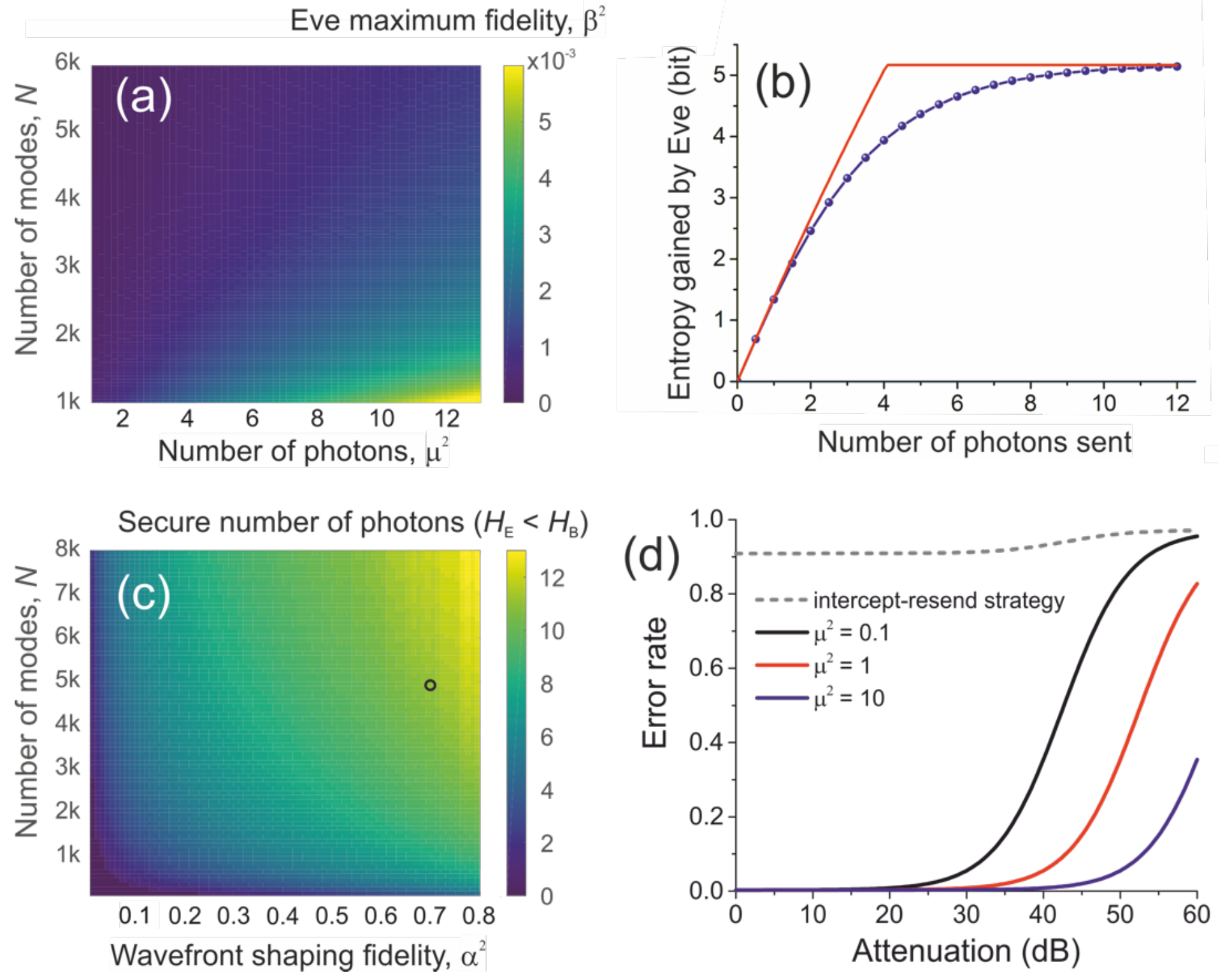

Fig. 3. (a) The fundamental limit of the best possible fidelity $\beta^2$ with which Eve can measure the wavefront with a low photon budget as a function of the number of photons, $\mu^2$ and the number of modes, $N$. (b) Entropy per transmitted symbol gained by Eve for $N = 1500$ fiber modes and $S = 36$ symbols. The connected blue dots represent the exact entropy function (evaluated through numerical integration) and the red line is a closed form expression that gives a simple upper limit (see Appendix B for details). The result is independent of the wavefront shaping fidelity. (c) The maximum number of photons per symbol measured by Eve that guaranties security ($H_E < H_B$) as a function of wavefront shaping fidelity, $\alpha^2$ and the number of modes $N$, for an alphabet consisting of $S = 36$ symbols. (d) The fractional secure qudit error rate (error per detected symbol) plotted as a function of the attenuation $a$, for $N = 5000$ modes, wavefront shaping fidelity $\alpha^2 = 0.7$ (the small circle in (c)), and the average number of emitted photons $\mu^2 = 0.1$ (the black solid line), 1 (the red solid line), and 10 (the blue solid line). An intercept-resend attack on all photons induces an error rate of at least 0.9 (grey dashed line).

Now we consider a photon-number splitting attack. We assume that Eve can perfectly measure a transmission matrix of the whole fiber except the scrambled parts of Alice and Bob and can build an ideal lossless high-dimensional spatial light modulator and lossless channel. In that case Eve can in principle choose to measure only a fraction of the photons and propagate the unmeasured photons to Bob's scrambler without losses. We demonstrated that Eve needs to measure a relatively large number of photons to get the same information as Bob. To keep the protocol secure against a photon-number splitting attack, the total number of photons per symbol sent by Alice should be less than the secure number of photons in Fig.3(c).

The success of more general attacks based on, e.g., optimal cloners, are bound by the general result from Bruss and Machiavello [38] that the fidelity of any estimate of a $d$-dimensional quantum state is limited to $\beta^2 = (1 + n)/(n + d)$, where $n$ is the number of copies that is available of the pure quantum state under investigation. In our method, $n = \mu^2$ and $d = 2N$, taking into account $N$ spatial modes and two polarization states. This bound is only slightly higher than the result derived in [34] for phase measurements of weak coherent light. For $N = 1500$ and $\mu^2 = 10$, best possible fidelity with which Eve can measure the wavefront $\beta^2 = 0.0033$ and fidelity of optimal quantum state cloner is $\beta^2 = 0.0037$.

### *3.4 Security analysis for the calibration phase*

To ensure that Eve cannot intercept the calibration phase, we made two small modifications to the stepwise sequential wavefront shaping algorithm. First of all, Alice sends the wavefronts in a random order, so that Eve does not know what pattern was sent. Second, Alice sends spatially shaped waves that only contain a low number of photons at a time. Alice changes the phase of a single segment of the light modulator, meaning that the wavefront is largely plane. We can describe this wave as a superposition of a plane wave and a part corresponding to the modulated segment. Since the intensity is distributed equally over all $N_{\mathrm{seg}}$ segments, the value of $\mu^2$ used in the analysis (see Appendix B) would be $\mu^2 = \bar{n}_c/N_{\mathrm{seg}}$, with $\bar{n}_c$ the average number of photons send during a single pulse in the calibration measurement (in our case, amounting to $\mu^2 = (80 \pm 30)/1156 \approx 0.07$ photons). From this measurement Eve cannot determine what symbol was sent with any reasonable confidence: the best possible fidelity $\beta^2$ with which Eve can measure the symbol is less than $1/N$ meaning a random result. So, she cannot intercept the calibration phase and retrieve the transmission matrix. Note that the plane-wave component of the wavefront contains much more energy. However, this component does not carry any information, so Eve does not gain anything by intercepting it.

After all measurements, Alice sums all corresponding frames for each measurement. She now has high-contrast speckle images that she can use to calculate the transmission matrix. Alice and Bob now share a secret in a special way: Bob has a scrambled fiber configuration in physical form, but has no knowledge of its optical transfer function, whereas Alice knows the transmission matrix of that fiber. Eve on the other hand, does not know what frames belong to what measurement, so she has no way to combine the frames into a high-quality image. In addition, if Eve intercepted the communication during the calibration phase, she will have sent arbitrary wavefronts to Bob, thereby breaking the correlation between Alice's transmitted field and Bob's response. As a result, when Alice now sums the corresponding frames, she will get low-contrast (uniformly distributed) images. This way, Alice can detect the presence of Eve without giving away any information. To ensure that Eve cannot launch a passive attack and use light reflected from the fiber facet or optics to measure any part of the transmission matrix of the Bob's scrambling fiber, Alice and Bob should install optical isolators.

## 4. DISCUSSION

In our experiments, the bit rate was limited by the speed of the camera. With high-speed cameras or avalanche photodiode arrays, the limitation of the proposed method might be the speed at which phase masks can be changed. Specialized hardware can be used to encode patterns with a speed as high as 350 kHz [39]. Together with the high dimensionality of the used alphabet ($H_B = \log_2 36 = 5.2$ bit per pulse) and unity wavefront shaping fidelity, it gives rise to a secure bit rate of up to 1.8 Mb/s. The bit rate limit can be increased twofold without any changes in the setup and equipment by increasing the number of symbols [40] and the number of guided modes of the multimode fiber. Commercially available multimode fibers support up to $3\cdot10^6$ modes, and allow to increase the dimensionality of the quantum channel and consequently the level of security.

Losses and instability of the fiber over time may influence the performance of the proposed approach. In the case of a *weak coherent light source*, losses affect the security of a protocol due to the possibility of a photon-number splitting attack. We assume that Eve can replace the lossy communication channel with a lossless one and capture all photons that are originally lost. Our analysis shows that for a wavefront shaping fidelity $\alpha^2 = 0.7$, number of modes $N = 5000$, and the average number of photons $\mu^2 = 1$, the secure qudit error rate is 12% (see Fig. 3(d)) for more than 40 dB of total attenuation, whereas Eve needs to intercept at least $\mu^2 = 10$ photons to get the information as presented in Fig. 3(c). Further improvements could be achieved by implementing the concept of decoy states. In our lab conditions, multimode fiber remains stable overall many days, but real-life implementation may require frequent recalibration due to high sensitivity of MM fiber to temperature and mechanical stress. Instability over time provides automatic rotation of the physical key.

Because of the unique properties of the method we introduce here, data is transferred in a deterministic manner, allowing Bob to decode the information instantaneously during the communication. Although this potentially allows to directly communicate secretly [41], there is always the risk of leaking a fraction of this information to Eve and we do not follow that route here and instead assume that our method is used to establish a secret key.

To summarize, we propose and implement a new type of quantum key exchange method by encoding information into a few-photon light pulse decomposed over guided modes of a multimode fiber. The new method is based on the difficulty to measure a quantum state without knowing the basis in which the state is diagonal combined with random light scrambling and secure wavefront shaping. We showed the security of the method against standard intercept-resend and beam-splitting attacks in case of a single-photon source, as well as in the case of a weak coherent light source with a low (relative to the number of modes) photon number. Moreover, it works with a high-dimensional alphabet, and can be naturally extended to larger distances.

## Appendix A: Details of the experimental setup

Alice uses the continuous-wave linearly polarized output of a He-Ne laser with a wavelength of 633 nm (see Fig. 1 'Alice'). Neutral-density filters in combination with a half-wave plate and a polarizing beam splitter are used to reduce the power to the desired level. A single-mode fiber is used to clean the laser mode and to expand the laser beam in order to match the surface of a spatial light modulator. To control the spatial phase of the coupled light with high speed, Alice uses a 1920x1200 Vialux V4100 digital micromirror device (DMD). The lenses $L_1$ and $L_2$ are placed in a 4f-configuration to image the phase mask on the back focal plane of a coupling objective. A pinhole in the Fourier plane blocks all the diffraction orders except the 1st, which encodes the desired spatial phase distribution. An objective with NA = 0.4 (Olympus) is used to couple light into a conventional step-index multimode fiber (Thorlabs, FG050UGA) with a silica core of $d_c = 50$ μm diameter and numerical aperture NA = 0.22. Such a fiber has a normalized frequency $V = \pi\, d_c\, \mathrm{NA}/\lambda \approx 55$ for our laser wavelength, $\lambda = 633$ nm, and sustains approximately $N = 1500$ guided modes [16]. The part of the experimental setup described above belongs to Alice and potentially allows encoding the desired symbol with a speed of up to 22 kHz. The multimode fiber serves as a quantum channel to deliver symbols to Bob in such a way that anyone who wants to listen to the communication at any arbitrary position along the fiber fails. We mimic the situation when Eve has access to the fiber without Alice and Bob knowing so, dividing the quantum channel into two parts. Each piece of the multimode fiber is 3 meters long. Eve's part of the setup includes two objectives (Olympus, NA = 0.4) served to close a gap between two pieces of fiber and to image the fiber output on her camera (see Fig. 1 'Eve'). Bob's part of the setup includes one objective (Olympus, NA = 0.4) to image the fiber output on the camera (see Fig. 1 'Bob'). Bob and Eve use a high-sensitive intensified charge-coupled device (ICCD) camera to detect light at the single-photon level (see 'Image processing' section). Light will pass from Alice to Eve, where it is split at M2, from where a part falls on Eve's camera and a part continues to Bob through the second fiber. During the calibration and characterization of Bob's part we use the first pathway and during the characterization of Eve's part we use the second pathway.

*Calibration and wavefront-shaping procedure*

The complex wavefront-shaping algorithm to create a focused laser spot on the fiber output facet without information about the configuration of the fiber-based scrambler is as follows. We use the DMD to control the spatial phase profile of light at the fiber output facet. Each mirror of the DMD can be set to two different tilt angles. By controlling the tilt of every mirror, 2D binary gratings can be created [42]. With an appropriately tilted input light field, the diffracted light propagates along the normal of the DMD surface (see Fig. 1 'Alice'). The DMD area of $510 \times 510$ pixels covered by an expanded laser beam is divided into $N_{\text{seg}} = 34 \times 34$ segments. Each segment consists of $15 \times 15$ micromirrors. Alice modulates the phase of a single segment by shifting the grating pattern over $2\pi$ in three steps. In total 3468 different phase masks were used. To guarantee the security of the wavefront shaping procedure, each pulse Alice sends contains only $80 \pm 30$ photons. The single segment that is modified contains on average less than a photon. Bob measures the intensity in $S = 36$ points on the fiber output facet or on a camera frame and sends this information back to Alice via a classical channel. In our experiments, a grid of $6 \times 6$ points with 3.2 µm step size on the fiber output is used. However, the position of the symbols is not restricted and can be selected randomly. They repeat the measurements 50 times for each phase mask in random order. After all measurements, Alice sums all corresponding frames for each measurement. She now has high-quality images that she can use to calculate the transmission matrix. The time required for the optimization procedure is 3 minutes and limited by the frame rate of the camera we used.

The security of the wavefront-shaping procedure is based on the same arguments as that of the security of the quantum communication after completion of the wavefront shaping: the no-cloning theorem forbids an attacker to fully characterize the light pulse containing fewer photons than the number of fiber modes and scrambled by the multimode fiber. Eve can record the light that is sent by Alice. However, since there is very little difference between the different phase masks (since there is less than one photon in each segment), it is impossible for her to know what segment Alice is probing. Therefore, if Eve 'replays' the field to Bob's fiber, she will just have to send some arbitrary wavefront. Now, when Alice sums Bob's camera frames, she will not get a high-quality speckle image. Instead, she will get a completely washed-out noise pattern, which will signal the presence of the eavesdropper. Since Alice and Bob have not shared any secrets yet, no information can have leaked.

We have characterized the fidelity (efficiency) of the wavefront shaping by the parameter $\alpha^2 = P_f / P_0$, where $P_f$ is the power in the focus area with a center corresponding to that of the focal spot and a diameter equal to the FWHM of the Gaussian spot. $P_0$ is the total power on the fiber output. We assume that any losses don't influence the wavefront shaping quality but only the total signal. In the presented experiments, $\alpha^2$ is 10%. In our experimental setup, the fidelity was limited by Eve's interception part in the transmission line (see Figure 1). In practice, an uninterrupted fiber should be used. For such fibers, a fidelity close to $\pi/4$ was reported experimentally for phase-only wavefront shaping [23].

*Image processing*

The signal was recorded by a HiCAM 5000 High-speed Intensified Camera (Lambert Instruments, the Netherlands) with 512x512 pixels and a speed of 5000 fps. To keep well-defined sensitivity, the incident photon flux is restricted to not exceed 5 detection events per frame on average. To investigate the parameters of secure communication with a higher number of photons, we summed up the required number of frames measured in the same conditions. As a result, an event with two or more photons at one point can be easily detected.

## Appendix B: Security

*Attacker model*

We assume that Bob has $S$ detectors, where $S < N$. We here aim at proving security of our multimode-fiber-based method of communication for certain attacker models, giving security bounds. We consider the following classification of attacker models, following Scarani *et al*. [4]

We analyze *individual attacks* and their subfamily of *intercept-resend attacks* where Eve chooses one basis in which to measure. As the name indicates, in this class of attacks, Eve intercepts the signal somewhere on its way from Alice to Bob, performs a measurement on it, and prepares a new signal that she sends to Bob (see Fig. 1). The part of the fiber controlled by Bob is inaccessible to Eve. We also consider a *photon-number splitting attack*.

*Collective attacks*, which imply that Eve keeps data in a quantum memory until the end of the classical post processing, are, fortunately, not applicable for our method. The great advantage of our multimode-fiber-based method is that it doesn't require a quantum key sifting step or any other classical post processing. Eve would not benefit from storing all the intercepted light in a quantum memory during the calibration phase because 1) she will never find out how to combine them successfully (only Alice knows the order of the phase masks) and 2) her intercept attack would be detected, since Eve would be forced to send random wavefronts to Bob.

*General attacks*, includes many possible variations and cannot be efficiently parametrized. Nevertheless, bounds for unconditional security have been found in many cases and in all these cases, it turns out that the bound is the same as for collective attacks [4].

*Hacking attacks* are related to the weaknesses of a practical implementation. The best-known example is the family of *Trojan horse attacks*, in which Eve probes the settings of Alice's and/or Bob's devices by sending some light into the system and register the reflected signal. However, in the setup where light goes only one way, the solution against Trojan horse attacks consists in using an optical isolator [4], which can be easily implemented in our setup.

We consider two main scenarios:

1) Alice uses a single-photon source. In this case, Eve is restricted to intensity measurements only.
2) Alice uses a weak coherent light source. In that case, we assume that Eve is able to record the field in each of the fiber modes independently.

In both scenarios, we do not specify the basis in which Eve does the measurements, but assume that Eve chooses one basis and does not adapt her basis choice. This situation applies when, e.g., Eve images a tapped MMF intensity profile with a camera. We also assume that Eve cannot copy and physically reproduce all essential features of the fiber between Eve and Bob in some passive optical system that would focus each symbol onto a different detector because Eve cannot directly measure the transmission matrix of the scramblers of Alice and Bob and cannot reconstruct the transmission matrix from intercepted optical signals.

*Information gained by Bob*

The maximum amount of information that Bob can read out per received photon is $H_B = \log_2(S)$. In case of $S = 36$ symbols this maximum is $H_B \approx 5.2$ bit of information. However, for a real-life situation, the fidelity of wavefront shaping will not be unity. The information Bob can gain per single-photon detection event in this scenario can be calculated as

$$H_B \equiv H(B) - H(B|s), \tag{B1}$$

where $H(B)$ is the entropy of the received alphabet, $H(B|s)$ the conditional entropy at the receiver side under the condition that symbol $s$ was sent. We consider the situation when Bob takes into account only the symbols for which he got a click on one of the $S$ detectors. The probability of getting a click on detector $b$ in a case a random unknown symbol is sent, is

$$P(b) = \frac{1}{S}. \tag{B2}$$

The total entropy, given that there was a click on one of the detectors is

$$H(B) = -\sum_{b=1}^{S} P(b)\log_2 P(b) = \log_2(S). \tag{B3}$$

Now we calculate the conditional entropy $H(B|s)$, which is the amount of information that is needed to describe measurement outcome $B$ given that the symbol $s$ is known. The probability of getting a click on detector $b$ given that symbol $s$ was sent, is

$$P(b|s) = \begin{cases} \dfrac{\alpha^2}{\alpha^2 + \dfrac{1-\alpha^2}{N-1}(S-1)}, & \text{if } s = b \\[2ex] \dfrac{1-\alpha^2}{(N-1)\left[\alpha^2 + \dfrac{1-\alpha^2}{N-1}(S-1)\right]}, & \text{if } s \neq b \end{cases} \tag{B4}$$

where $N$ is the number of modes, $S$ is the number of symbols, and $\alpha^2$ is the fidelity of the wavefront shaping. Here we consider only the events with a click on a detector. As a result, the conditional entropy is

$$\begin{aligned} H(B|S) = -(S-1)\,P(b|s \neq b)\log_2\big(P(b|s \neq b)\big) \\ - P(b|s = b)\log_2\big(P(b|s = b)\big). \end{aligned} \tag{B5}$$

Using equations (B1), (B3) and (B5) we find that the information, which Bob can get per single-photon detection in case of non-perfect wavefront shaping, is

$$\begin{aligned} H_B = \log_2(S) + (S-1)\,P(b|s \neq b)\log_2\big(P(b|s \neq b)\big) \\ + P(b|s = b)\log_2\big(P(b|s = b)\big). \end{aligned} \tag{B6}$$

This value was used to analyse the parameters for which the amount of information per photon gained by Bob in imperfect conditions becomes more than the theoretically possible maximum of information gained by Eve.

We also calculate the probability of getting a click on detector $s$ given that symbol $s$ was sent in case of non-ideal detectors and losses. Alice sends a symbol $s$ with field amplitude $\mu$. Attenuation of the system is $a$ (in dB). The dark count rate of Bob's detectors is $p_{\mathrm{dark}}$. Similar to Eq. B4 we can get the practical probability to detect the correct symbol as a function of the wavefront shaping fidelity $\alpha^2$ (the probability of getting a click on detector $b$ given that symbol $s = b$ was sent):

$$P_{\mathrm{pr}}(b|s = b, \alpha^2) = \frac{\alpha^2\mu^2 10^{-\frac{a}{10}} + p_{\mathrm{dark}}}{\alpha^2\mu^2 10^{-\frac{a}{10}} + \dfrac{1-\alpha^2}{N-1}(S-1)\mu^2 10^{-\frac{a}{10}} + S p_{\mathrm{dark}}}. \tag{B7}$$

As a result, the acceptable level of the error rate in the quantum channel between Alice and Bob can be calculated as

$$QER_{\mathrm{secure}} = 1 - P_{\mathrm{pr}}(b|s = b, \alpha^2). \tag{B8}$$

An eavesdropper employing an intercept-resend strategy on all photons would induce an additional qudit error rate. The error rate in the quantum channel can be used to detect the presence of Eve. We can calculate qudit error rate after Eve's interception as [33]:

$$QER_{\text{interception}} = 1 - P_{\text{pr}}(b|s = b, \beta^2\alpha^2\,), \quad \text{(B9)}$$

where $\beta^2$ is Eve's best possible fidelity: $\beta^2 = \mu^2/(\mu^2 + 2N)$ (see Ref. [34] for details).

*Security analysis for a single-photon Fock state*

We assume that Alice uses a perfect single-photon source to send symbols and that Eve can read the signal somewhere in the middle of the fiber, potentially close to Alice or close to Bob. Additionally, since Alice sends a single photon, we assume that Eve performs intensity measurements. We now find a strict upper limit of information that can be gained by Eve by considering the hypothetical scenario where Eve somehow knows the basis (what field arrives at pixel $e$ of her detector when Alice sends a given symbol $s$), but is not able to build a passive linear-optical mode convertor to sort the intercepted photons efficiently into $S$ detectors. This scenario is excluded by the setup and our calibration process and considering it here is only a computational trick that allows us to estimate a strict limit.

Let's call this the field transmission function $t_{es}$. Neglecting losses and assuming Eve has a perfect detector (with $N$ pixels, one for each mode in the fiber), she will get exactly one click on one pixel of her detector for each symbol that is sent. We are now interested in calculating how much information Eve obtains by recording this single photon. The maximum amount of information that Eve can get is

$$H_E \equiv H(E) - H(E|s). \quad \text{(B10)}$$

Here, the entropy $H(E)$ is the amount of information that is needed to describe a measurement outcome $E$ and the conditional entropy $H(E|s)$ is the amount of information that is needed to describe this outcome when it is known which symbol $s$ was sent. By Bayes' rule for conditional entropy, the difference between the two entropies $H_E$ is the maximum amount of information that Eve can possibly gain. We are now left with the task of calculating the conditional entropy given that the symbol $s$ is known.

$$H(E|s) = -\sum_{e}^{N} P(e|s)\log_2 P(e|s)\,, \quad \text{(B11)}$$

with $P(e|s) \equiv |t_{es}|^2$ the probability for a photon sent as symbol $s$ to arrive at pixel $e$. Since there are no losses, $\sum_{e}^{N}|t_{es}|^2 = 1$. The exact value of $H(E|s)$ depends on the unknown, random transmission matrix elements and, therefore, is impossible to predict in advance. However, we can readily find the ensemble averaged value $\langle H(E|s)\rangle$, i.e. the conditional entropy averaged over all possible random transmission matrices of the fiber. To do so, we assume that $\langle|t_{es}|^2\rangle = 1/N$ and that the elements $|t_{es}|^2$ are drawn from independent exponential distributions. In the case of a large number of modes $N$, the distribution of $|t_{es}|^2$ equals $P(|t_{es}|^2) = N\exp(-|t_{es}|^2N)$. To calculate the expected value for $H(E|s)$, we average over realizations of disorder. Substituting $y_{es} \equiv |t_{es}|^2N$ we can write

$$\langle H(E|s)\rangle \equiv -\frac{1}{N}\sum_{e}^{N}\int_0^\infty \exp(-y_{es})y_{es}\log_2\left(\frac{y_{es}}{N}\right)dy_{es}. \quad \text{(B12)}$$

Since the distribution of $y$ does not depend on $s$ or $e$, we can omit averaging over all symbols

$$\begin{aligned}\langle H(E|s)\rangle &= -\int_0^\infty \exp(-y)\,y(\log_2 y - \log_2 N)\,dy \\ &= \log_2 N - \int_0^\infty \exp(-y)\,y\log_2 y\,dy.\end{aligned} \quad \text{(B13)}$$

This integral evaluates to

$$\langle H(E|s)\rangle = \log_2 N - \frac{1-\gamma}{\ln 2}, \tag{B14}$$

with the Euler constant $\gamma \approx 0.577216$. Using (B1) we finally find

$$\langle H_E\rangle = \frac{1-\gamma}{\ln 2} \approx 0.61 \text{ bit}. \tag{B15}$$

Hence, Eve gains only 0.61 bit of information per transmitted symbol at best, regardless of the number of symbols or modes even in the extreme case when she knows the exact basis. Under the same conditions, Bob will have at maximum $H_B(S) = \log 2(S) \approx 5.2$ bit of information per transmitted symbol for 36 symbols. As a result, $\langle H_E\rangle << H_B$, guaranteeing the security of the method.

*Security analysis for coherent state and phase measurements*

We assume that Eve is somehow able to record the field in each of the fiber modes independently, e.g. by a homodyning technique. The accuracy of these measurements will be limited by shot noise in the reference. When Alice sends a symbol $s$ with field amplitude $\mu$, at Eve's side the field in mode $e$ equals $E_e = \mu t_{es}$. However, due to shot noise, Eve will actually measure $(\mathbf{E}_{\text{Eve}})_e = \mu t_{es} + \xi_e$, with $\xi_e$ a noise term with $\overline{\xi_e} = 0$ and $\overline{|\xi_e|^2} = 1$ photon, where $\overline{\cdot}$ indicates averaging over measurements [34].

We now proceed to calculate the amount of information that Eve gains by doing this measurement. Eve can first project her measurement of the field at the fiber output $\mathbf{E}_{\text{Eve}}$on the basis of symbols by simply multiplying with $t_{es}^{-1}$. When all modes of the fiber are used ($S = N$) and the fiber is lossless, $t_{es}$ is unitary. After performing the transformation, Eve will have found the vector $\mathbf{E} \equiv t_{es}^{-1}\mathbf{E}_{\text{Eve}}$, where each element in $\mathbf{E}$ corresponds to a symbol. Averaged over measurements we find $\overline{\mathbf{E}} = \mu_s$, where $\mu_s$ is a vector with a value of $\mu$ at index $s$, and zero everywhere else. Importantly, since the transformation is unitary and the noise in different components of $\mathbf{E}_{\text{Eve}}$ is uncorrelated, the transformation does not alter the noise statistics of the elements of $\mathbf{E}$: i.e. the noise term has a complex Gaussian distribution with $\overline{|\xi_e|^2} = 1$ photon. When $S < N$, it is still possible to define a unitary matrix $t_{es}$ that maps all possible symbols to the first $S$ indices of $E$ and maps all unused state to the remaining indexes. Eve can simply discard the values at these remaining indices since they will never contain any information. Therefore, below we consider $\mathbf{E}$ to be a vector of length $S$.

When Alice sends symbol $s$, the transformed vector $\mathbf{E}$ is drawn from the probability density function

$$P(\mathbf{E}|s) = \frac{e^{-\|\mathbf{E}-\mu_s\|^2}}{\pi^S}. \tag{B16}$$

The differential entropy of this complex multivariate normal distribution is given by $H(E|s) = S\log_2\pi + S/\ln(2)$. We need to compare this value to the entropy for the case that it is not known what symbol was sent. In this case, the probability density equals

$$P(\mathbf{E}) = \frac{1}{S}\sum_{s=1}^{S}\frac{e^{-\|\mathbf{E}-\mu_s\|^2}}{\pi^S}. \tag{B17}$$

Unfortunately, there is no known closed-form expression for the entropy of such a Gaussian mixture [43]. Therefore, we use a Monte Carlo approach to calculate the entropy of (Eq. B10) numerically (see Figure 3(b)). Additionally, it is straightforward to derive an upper limit for the entropy. First, we realize that the real parts of the components of $\mathbf{E}$ all have a mean value of $\mu$/S and a variance of $\frac{1}{2} + \frac{(S-1)}{S}\left(\frac{\mu}{S}\right)^2 + \frac{1}{S}\left(\mu - \frac{\mu}{S}\right)^2 = \frac{1}{2} + \frac{\mu^2(S-1)}{S^2}$. The imaginary parts have a mean of 0 and a variance of 1/2. The maximum entropy distribution for a given variance is Gaussian. Therefore, our distribution (see Eq. B17) must have a lower entropy than a Gaussian with the same variance. The differential entropy for this Gaussian is simply given by $H(\boldsymbol{E})_{\text{upper}} = ½\log_2(|2\pi e\Sigma|) = S\log_2\pi + S/\ln(2) + (S/2)\log_2(1 + 2\mu^2(S-1)/S^2)$ with $\Sigma$ the covariance matrix, which is diagonal in our case. Half of the diagonal elements are ½ + $\mu^2(S-1)/S^2$ (for the real parts), the rest (for the imaginary parts) is ½, so $|2\pi e\Sigma| = (\pi e)^S\left(\pi e(1 + 2\mu^2(S-1)\,S^{-2})\right)^S$. The upper limit for the entropy gained by Eve follows by subtracting $H(E|s)$, and also realizing that Eve cannot collect more information than what is being sent by Alice.

$$H_E < \min\left[\frac{S}{2}\log_2\left(1+2\mu^2\frac{S-1}{S^2}\right), \log_2 S\right]. \tag{B18}$$

As long as $H_E < H_B$, the method remains secure even in the extreme case of a weak coherent light source with more than 1 photon per pulse in a non-secret basis.

## Funding

Nederlandse Organisatie voor Wetenschappelijk Onderzoek (QuantERA project QUOMPLEX grant number 680-91-037, NWA Startup Quantum NanoKeys grant 40017607, Veni grant 15872, Vici grant 680-47-614); European Research Council (Starting grant 678919).

## Acknowledgments

We thank Klaus Boller, Ad Lagendijk, Mehul Malik, Ravi Uppu, and Willem Vos for discussions.

## Disclosures

The authors declare no conflicts of interest.